\documentclass[aps,prl,twocolumn,showpacs,groupedaddress,a4paper]{revtex4}  
\usepackage{graphicx}  
\usepackage{bm}        
\usepackage{amssymb}   
\usepackage{color}

\begin{document}

\title{From single particle to superfluid excitations in a dissipative polariton gas}

\date{\today}

\author{V. Kohnle$^{1^*}$, Y. L\'eger$^1$, M. Wouters$^2$, M. Richard$^3$, M. T. Portella-Oberli$^1$, B. Deveaud-Pl\'edran$^1$}

\affiliation{
$^1$Institute of Condensed Matter Physics, {\'E}cole Polytechnique F{\'e}d{\'e}rale de Lausanne (EPFL)\\
$^2$Institute of Theoretical Physics, Ecole Polytechnique F\'ed\'erale de Lausanne, CH-1015 Lausanne, Switzerland\\
$^3$Nano-Physics and Semiconductor group, N{\'e}el Institute, CNRS Grenoble, France}

\begin{abstract}
Using angle-resolved heterodyne four-wave-mixing technique, we probe the low momentum excitation spectrum of a coherent polariton
gas. The experimental results are well captured by the Bogoliubov transformation which describes the transition from single particle excitations of a normal fluid to sound-wave-like excitations of a superfluid. In a dense coherent polariton gas, we find all the characteristics of a Bogoliubov transformation, i.e. the positive and negative energy branch with respect to the polariton gas energy at rest, sound-wave-like shapes for the excitations dispersion, intensity and linewidth ratio between the two branches in agreement with the theory. The influence of the non-equilibrium character of the polariton gas is shown by a careful analysis of its dispersion. 
\end{abstract}


\maketitle 

\begin{figure}
		\includegraphics[width=0.48\textwidth]{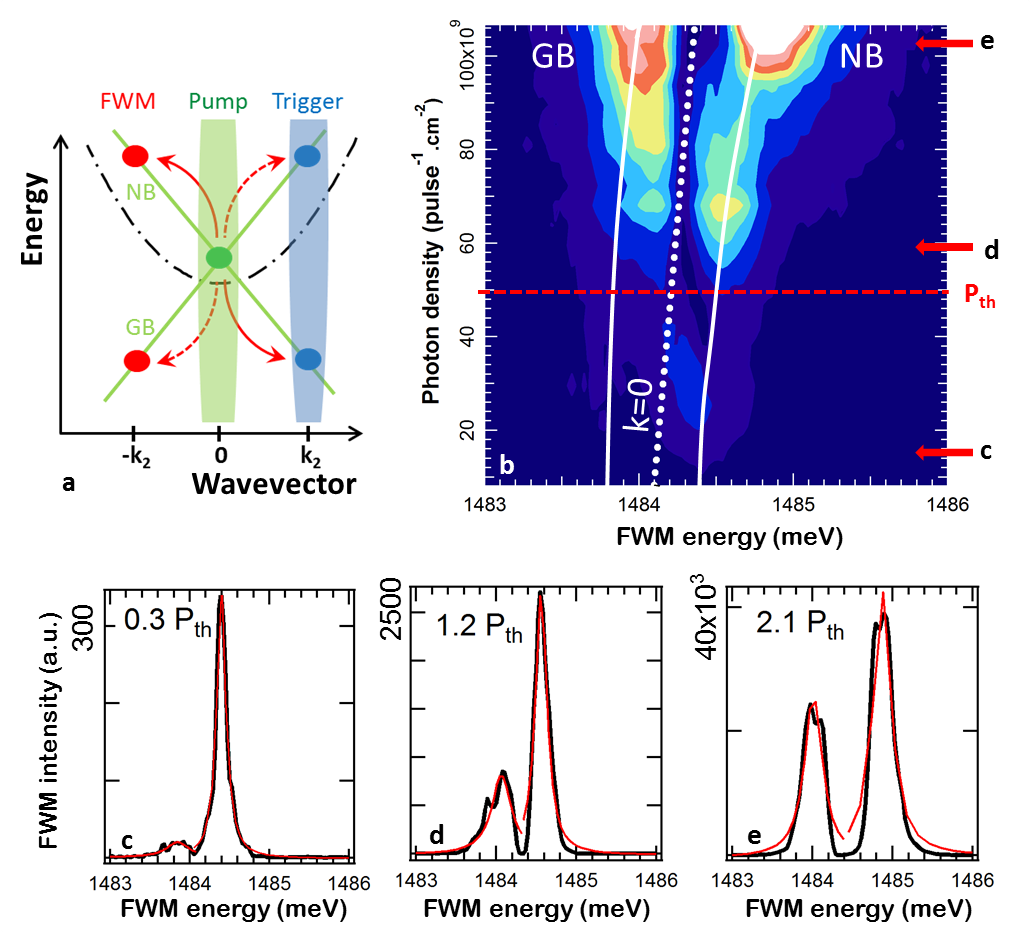}
	\caption{\textbf{a} With a fs-pump pulse the coherent polariton gas is created.	The trigger pulse gives rise to two distinct parametric scatterings (dashed or plain red arrows). 
\textbf{b} FWM spectrum (linear scale) as a function of the excitation photon density for a trigger wavevector k=1 $\mu m^{-1}$. The white dashed line shows the k=0 polariton energy and the two white solid lines guide the eye and display the evolution of the NB and GB. A threshold at $P_{th}$= $5.10^{10}$ photons.pulse$^{-1}cm^{-2}$ is observed indicated by the red dotted line. \textbf{c}, \textbf{d} and \textbf{e} show FWM spectra in the perturbative regime, short above threshold and the superfluid regimes respectively. The red curves display the Lorentz fits for the two resonances. \label{fig:power}}	
\end{figure}

\begin{figure}
		\includegraphics[width=0.48\textwidth]{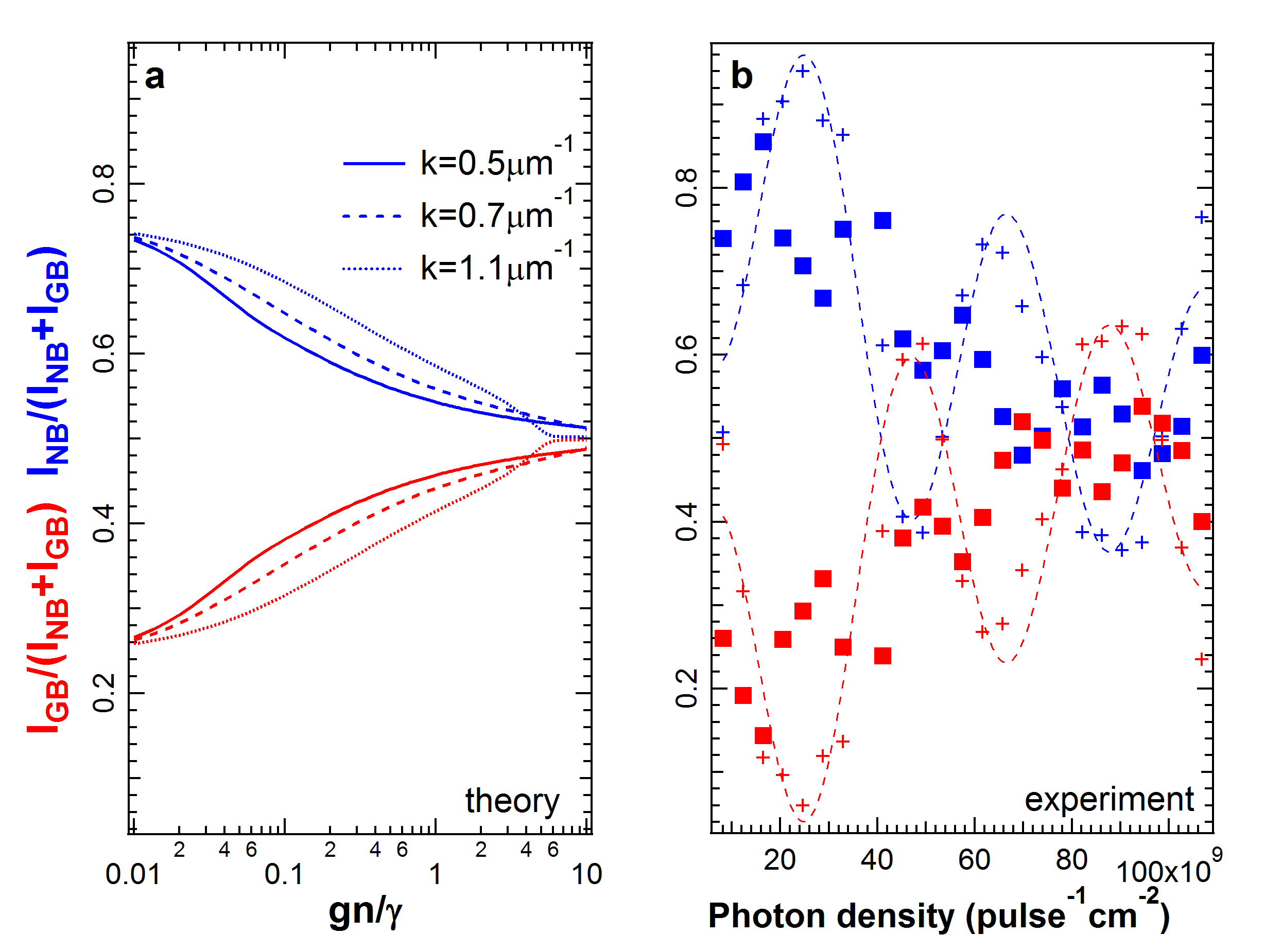}
	\caption{\textbf{a} Calculated intensity evolution of the NB (blue) and GB (red) emission contribution plotted for different perturbation wavevectors. Calculation parameters: $\hbar \gamma = 0.1 meV$, $\hbar/2m =0.5 meV. \mu m^2$. \textbf{b} Normalized intensity of the NB and GB (blue and red crosses respectively and dashed lines for the fits). The blue and red squares display the normalized intensities of the NB and GB with removed oscillating part.  
\label{fig:Theory}}	
\end{figure}

The demonstration of superfluidity of Helium II by Kapitsa \cite{Kapitza}, Allen and Misener \cite{Al} in 1938 triggered fundamental theoretical research for understanding quantum fluids. London quickly linked superfluidity with Bose Einstein condensation\cite{London}, stressing the importance of the bosonic nature of the particles in the phenomenon. Meanwhile, Landau developed the idea of sound-like excitations of a superfluid\cite{Landau}. These intuitions were later confirmed by Bogoliubov whose microscopic theory of the weakly interacting Bose gas\cite{Bog} reveals a superfluid phase in which elementary excitation is a coherent superposition of counter-propagating particle and hole.\\
In a superfluid at equilibrium, the momentum dispersion of the excitations should deviate substantially from the parabolic single particle case. In addition to a global blueshift due to repulsive particle-particle interactions, the dispersion turns linear for small wavevectors as a result of the sound-wave nature of the superfluid elementary excitations. In addition to the normal positive energy branch (NB), the Bogoliubov dispersion is expected to display a negative energy "ghost" branch (GB), resulting from the hole component of the excitation \cite{Pit}. The NB and GB are the mirror image of each other and represent the dispersion of Bogoliubov excitations \cite{Wouters1}. Superfluid excitations have been extensively investigated in both He superfluids \cite{Jack, Woods,Hart} and ultracold atomic gases \cite{Oz, stein, Jin}. First signs of a Bogoliubov transformation have been given by Vogels et al \cite{Ketterle} in atom condensates.\\
In a semiconductor microcavity where quantum well excitons and cavity photons are strongly coupled, light-matter quasi-particles called polaritons are formed. These composite bosons show unique properties which promote them as a model system of the weakly interacting Bose gas.
Indeed, since demonstration of Bose-Einstein condensation of polaritons \cite{Kasprzak2006}, superfluidity of coherent polariton fluids has been assessed the observation of zero-viscosity and critical velocity by Amo et al \cite{Amo, Amo2}. Very recently superfluid excitations have been investigated in a nonresonant polariton condensate \cite{Yam}. The negative Bogoliubov dispersion branch could surprisingly not be observed though.\\
However, the peculiarity of polaritons with respect to other superfluids is their non equilibrium nature. In the case of non-resonant excitation of the polariton condensate the theory predicts the appearance of a flat diffusive region in the Bogoliubov dispersion around k=0, resulting from the coupling between the exciton reservoir and the polariton field \cite{Wouters1}. Under cw resonant excitation no exciton reservoir is created and the excitation spectrum is therefore much closer to the equilibrium Bogoliubov spectrum \cite{Car}. Under pulsed resonant excitation however, we expect strong modifications in the polariton fluid dispersion due to the decay of the polariton population.
\\
In the present work, following the suggestion of a recent theoretical work \cite{Wouters2}, we choose the following strategy to investigate the superfluid character of a dissipative coherent polariton gas (CPG). We use the four-wave mixing (FWM) technique to stimulate emission from both the normal and negative energy branch in the dispersion of a resonantly created CPG (Fig.\ref{fig:power}a). Our experimental access to both these dispersions allows us to observe the modification of the polariton gas excitations and to conclude on the superfluid character of the CPG, e.g. Bogoliubov transformation. Contrary to reference \cite{Wouters2} we use circularly polarized pulsed excitation and thus probe the excitations of a spin polarized CPG in the dissipative regime. We can analyze the dynamics of the CPG excitations as the polariton density decays. We evidence the influence of dissipation on the excitation spectrum of a CPG resulting in an asymmetry of the NB and GB in the dispersion and an overestimated sound velocity.\\
We investigated a single 8 nm $In_{0.04}Ga_{0.96}As$ quantum well embedded in a GaAs wedge-shaped-cavity \cite{Houdre}. The Rabi-splitting is 3.4 meV at zero detuning. The sample is kept in a continuous flow cryostat at around 5K. We use a two pulse heterodyne FWM setup. For optical excitation a 80 MHz pulsed Ti:Sapphire laser is used. 
The 250 fs pulses are centered on the lower polariton k=0 energy and cover the energy range of both lower and upper polariton dispersion curves. The laser beam is split in three parts: a reference, a pump(index 1) and a trigger (index 2). These last two co-circularly polarized beams are radio-frequency shifted with acousto-optic modulators (AOMs) by  $\omega_{1}$ (75 MHz) and   $\omega_{2}$ (79 MHz) and are focused on a $~ 100 \mu m$ diameter spot on the sample. 
The pump pulse generates a CPG at rest ($k_1$=0). In a second time the trigger pulse stimulates excitations in the CPG. It is ten times weaker than the pump pulse and carries a finite wavevector k, which is varied between $k_2=0$ to 1 $\mu m^{-1}$. Stimulated parametric scattering of polaritons from the CPG results in the generation of a FWM signal at opposite wavevector $-k_2$. The FWM signal is selected at $k_{FWM}$= $2k_1-k_2$=$-k_2$ and is directed into a mixing AOM together with the reference. The AOM, driven at  $\omega_{FWM}$=$2\omega_1- \omega_2$ (71MHz), up-shifts the diffracted reference field and down-shifts the diffracted signal field by $\omega_{FWM}$. This results in a spectral overlap of reference and signal field and a $\pi$-phase shift for the two heterodyne channels. The mixed beams are then dispersed in a spectrometer. To recover the interference term between the FWM signal and the reference field, we subtract the two  $\pi$-shifted interferograms. In this way the classical noise, e.g. photoluminescence, is largely suppressed. From the measured interference we retrieve then the signal in amplitude and phase by spectral interferometry \cite{Hall}.\\
Using nonlinear spectroscopy off-resonance polariton scattering has already been observed by Savvidis et al. \cite{Sav}. It resulted in multi-wave-mixing processes between pump, signal and idler modes of a polariton OPO, revealing non negligible nonlinearity in the microcavity. However, the experiment could not allow them to demonstrate any modification of the dispersion in the vicinity of the CPG momentum. In our case, thanks to the heterodyne detection scheme, we precisely obtain an experimental access to the low momentum excitations of the polariton fluid. Heterodyning allows us to separate the FWM signal from linear coherent emission and higher order nonlinear emission of the CPG when the standard angular selection becomes inefficient.\\
Bogoliubov transformation should result in a drastic change of the FWM response in terms of intensity contributions of the NB and GB modes as well as in terms of dispersion curvature. We therefore study the changes of the FWM spectrum with excitation intensity, and show that the nature of the CPG excitations changes from single particles to sound waves when increasing the polariton density. Any nonlinearity in the microcavity does result in the observation of a FWM signal on the negative energy side. Here we show that this "off resonant" emission changes drastically when the superfluid regime is reached. In this case the emission intensities of NB and GB become comparable.
\\
In Figure \ref{fig:power}b, we plot the measured FWM spectrum as a function of excitation density at a fixed pump-trigger delay of 6.5ps.  The trigger incidence angle was kept constant to stimulate excitations at wavevector 1 $\mu m^{-1}$ where the NB and GB are well separated. 
Two emission lines are observed. One lays below the k=0 polariton energy, the other one, laying above the k=0 energy, is the normal branch.
At low excitation intensities (below the threshold $P_{th}$) the emission from the NB is much stronger than the emission of the negative energy branch (Fig. \ref{fig:power}c). Besides the emission of the negative energy branch is symmetric to the NB emission with respect to k=0. These two features demonstrate that the system is still in the perturbative regime, where the dispersion is given by the parabolic single-particle one. The emission at negative energies is attributed to off-resonance polariton scattering processes. 
Experimentally we observe a threshold $P_{th}$ for the GB intensity at $5.10^{10}$ photons.pulse$^{-1}$cm$^{-2}$. 
Above $P_{th}$ (dashed line) the GB emission is steeply enhanced and becomes of the same order of magnitude as the NB emission. This shows that the excitations of the CPG have changed. Fig. \ref{fig:power}d shows the FWM spectrum just above $P_{th}$. Far above $P_{th}$ (Fig.\ref{fig:power}e), the average intensity ratio between GB and NB is around 0.9.
\\
The linewidth is also affected by the increase of the polariton density. In the low density regime the linewidth of the negative energy mode is much broader than the one of the NB mode. The Lorentz fits of the two resonances in Fig.\ref{fig:power}c (in red) exhibit a FWHM ratio between the GB and NB of 2.6 (FWHM: 0.14 meV(0.36 meV) for the NB (GB)). The change of the regime is accompanied with a redistribution of the linewidth. As its density increases and the ratio between the GB and NB intensity tends to one, the linewidths of the GB and NB become equal. Just above $P_{th}$ (Fig \ref{fig:power}d ) the FWHM ratio has already decreased to 1.8 (FWHM: 0.17 meV(0.30 meV) for the NB (GB)). At high density (Fig. \ref{fig:power}e) the NB and GB linewidths are 0.26 meV and 0.27 meV, respectively.
\\
The relative intensities of the positive and negative energy contributions of the FWM signal can be calculated in the lower polariton (LP) subspace, in which the polariton dynamics are described by the time-dependent Schr\"odinger equation:
\begin{eqnarray}
i\dot{\Psi}(\vec{r},t)=(\epsilon(\vec{\nabla})+g\left|\Psi(\vec{r},t)\right|^2-i\frac{\gamma}{2})\Psi(\vec{r},t)
+F_{p, t}(\vec{r},t)
 \label{schroedinger}
\end{eqnarray}
where $\Psi$ is the LP wavefunction, $\epsilon$ its kinetic enegy, $g$ the 
polariton interaction constant, and $F_{p ,t}$ the pump and trigger excitations. 
According to the design of our experiment, three modes must be considered. The CPG mode at k=0 is described by $\varphi_0$. In analogy with Bogoliubov's theory, we call u and v* the counter-propagating perturbative modes coupled by 
polariton interaction. The trigger pulse excites the v* mode and FWM is generated on the u mode.  The LP wave function reads as:
\begin{equation}
\Psi(\vec{r},t)=\varphi_0(\vec{r},t)+u(\vec{r},t)e^{i\vec{k}\vec{r}}+v^{*}(\vec{r},t)e^{-i\vec{k}\vec{r}}	
\end{equation}
Equation \ref{schroedinger} can be solved numerically and the spectrum of the u mode can be calculated to obtain the intensity ratio between GB and NB. Fig.\ref{fig:Theory}a provides the calculated density evolution for the emission contribution of NB and GB as a function of the normalized blueshift $(gn_0/\gamma)$ for different excitation wavevector. At low densities the calculated ratio is $1/3$ for any wavevector. As the density increases, the elementary excitations are modified and the u and v modes couple and redistribute polaritons in both positive and negative energy branch. The ratio tends faster to one for smaller k as expected by the equilibrium Bogoliubov theory, because for smaller k the transition is more Bogoliubov like.
The theoretical intensity ratio of 1/3 in the low density regime comes from different damping times of the FWM resonances: it is three times larger for the GB resonance than for the NB resonance. This is also what we can extract from the linewidth of the emission lines as shown in Fig. \ref{fig:power}c. These changes in the FWM spectrum demonstrate the appearance of the Bogoliubov excitations in the CPG.
\\
The excitation with fs-laser pulses generates populations of both LPs and upper polaritons (UPs). This yields oscillations in the excitation intensity spectrum for the NB as it can be seen in Fig \ref{fig:power}b. These oscillations complicate the discussion of the NB/GB intensity ratio. They could in principle be avoided by shaping the excitation pulses but this would not allow us reach the high intensities required for this experiment.
 The normalized intensities of the NB and GB are presented in Fig.\ref{fig:Theory}b as a function of excitation intensity. The oscillation imposed by the fs-pulses is clearly visible (crosses in \ref{fig:Theory}b). In order to suppress this feature due to the LP-UP mixing, we fitted the data with an combination of an exponentially decaying sine added to a standard exponential variation (dased lines in \ref{fig:Theory}b). The oscillating part is then subtracted from the data to obtain the mean GB (NB) intensity contribution (squares in \ref{fig:Theory}b). At low excitation power the NB emission is 4 times more intense than the GB emission. This is not far from the value of 3 expected by the pertubative theory. With increasing excitation power, the NB contribution decreases whereas the GB contribution increases. At high excitation power, the intensities of the two branches tend to the same value as it is expected for the Bogoliubov transformation.  The calculated density evolution for the NB and GB emission contribution (Fig.\ref{fig:Theory}a) is displaying the experimental behaviour.
 \begin{figure}
		\includegraphics[width=0.48\textwidth]{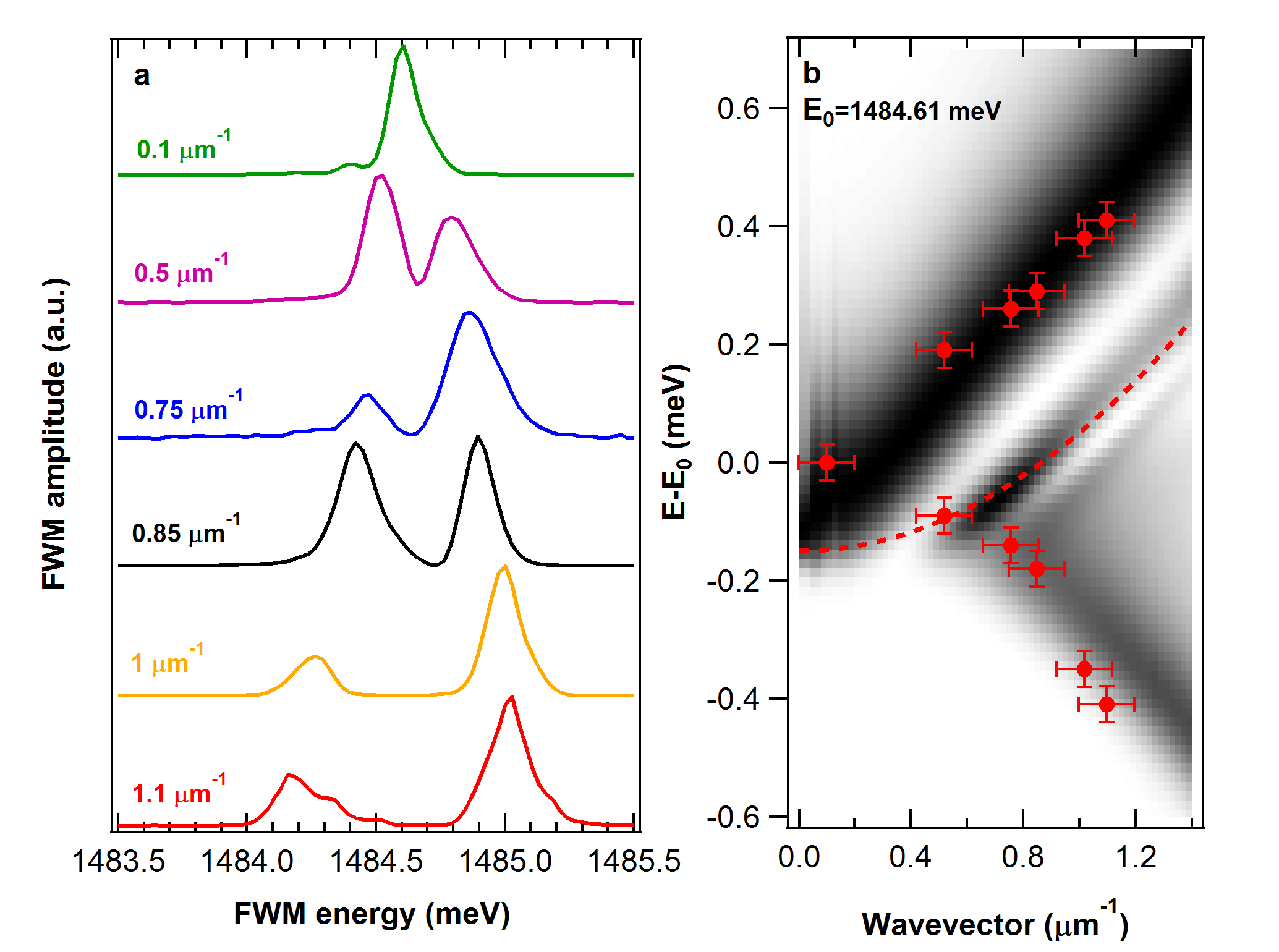}
	\caption{\textbf{a} FWM spectra at different excitation wavevectors. 
		\textbf{b} Energies of these lines (red dots) versus wavevector. We compare them with the single-particle parabolic dispersion (red-dashed line) and with the FWM signal calculated (black) using the Gross-Pitaevskii equations\cite{Car}. \label{fig:k}}	
\end{figure}
\\
To measure directly the curvatures of the NB and GB dispersion, we analyze the dependence of the FWM spectrum on the trigger wavevector. Figure \ref{fig:k}a displays the spectra of the FWM signals obtained for different wavevectors of the trigger pulse integrated between 5 and 6 ps delay. The intensity of the two pulses and quite importantly of the pump pulse (12.5 mW, $10^{11}$ photons/pulse/cm$^2$) was kept constant ensuring the same polariton density in the superfluid phase for all measurements. The energies of the two emission lines are plotted in Figure \ref{fig:k}b versus the trigger wavevector. The parabolic dispersion of polaritons obtained from transmission measurements in the low-density regime is also displayed for comparison (red dashed line in \ref{fig:k}b). Theoretically, in a steady state regime the dispersion of both the NB and GB is expected to be symmetric and linear \cite{Pit, Car}. Indeed the shape of the NB in our experiment is linear and strongly differs from the standard single-particle parabolic dispersion. This is the consequence of polariton-polariton interactions and a clear indication of superfluidity. However, there is no symmetry between the GB and the NB at low k.
Furthermore, if we extract a linear slope from the NB dispersion we find a sound velocity $c_s=0.6$ $\mu m/ps$. It does not correspond to the velocity obtained from the $k$=0 blueshift of 0.15 meV with the equilibrium formula: $c_s=\sqrt{gn/M}$.
The apparent sound velocity is thus larger than expected. This peculiar dispersion originates from the dissipative nature of polaritons. 
Taking the Bogoliubov equations, two physical processes can be pointed out: a blueshift of the whole dispersion corresponding to the blueshift of the condensate itself and the coupling of the two counter propagating modes of the Bogoliubov excitations leading to the change of curvature. If the former is linear with the particle density the latter is sublinear. The change of Bogoliubov-like curvature is therefore less sensitive to the polariton decay than dynamical blueshift. This leads to a steeper slope of the NB resulting in an apparent higher speed of sound and to the observed asymmetry in the time-integrated results.
A more complete theoretical study 
is provided in additional information.
\\ 
Our calculation reproduces properly the different features of the experimental dispersion: the linearity of the NB with an overestimated sound velocity and the asymmetry with the GB. The linearization of the Gross-Pitaevskii equation in the case of a decaying superfluid, assuming that the variations of the energies are slow, confirms the later result: the steeper slope of the NB originates from the time integration of the decaying polariton emission and not from a physical change of the sound velocity. Finally, one observes replicas of the NB at lower energy in the calculation which are also a result of the decaying superfluid density.\\
In conclusion we have evidenced the Bogoliubov transformation at the transition from a single particle gas to a superfluid in a semiconductor microcavity. We demonstrated the modification of the polariton dispersion curvature from parabolic in the pertubative regime to the peculiar linear dispersion of a dissipative superfluid. Despite their dissipative character, microcavity polariton gases can be qualitatively described in the frame of Bogoliubov theory of superfluids, closely linking polariton phenomenology to cold atom physics.\\
\indent We acknowledge R. Houdr\'e and U. Oesterle for providing the sample and SNSF for financial support.

\end{document}